\documentclass{jfm_arxiv}
\usepackage{commath}
\usepackage[nameinlink,noabbrev]{cleveref}%

\usepackage{graphicx}
\usepackage[numbers]{natbib}

\usepackage[locale=UK,
number-mode=math,
unit-mode=text,
per-mode=symbol,
number-unit-product=\ ,
retain-zero-exponent=false]{siunitx}[=v2]
\DeclareSIUnit{\pixel}{px}
\DeclareSIUnit{\fps}{fps}

\newcommand{\kindex}[2]{\ensuremath{{#1}_{\scalebox{0.65}{#2}}}}
\newcommand{\U}{U}
\newcommand{\Uinf}{\kindex{\U}{$\infty$}}
\newcommand{\Ma}{\mbox{Ma}}
\newcommand{\St}{\mbox{St}}
\newcommand{\subf}[1]{(#1)}

\graphicspath{{../figurematter/latex/figs/}}

\begin{document}
\title{All you need is time to generalise the Goman-Khrabrov dynamic stall model}

\author{Fatma Ayancik, Karen Mulleners\corresp{\email{karen.mulleners@epfl.ch}}}
\affiliation{Institute of Mechanical Engineering, \'Ecole polytechnique f\'ed\'erale de Lausanne, Switzerland}

\maketitle

\begin{abstract}
Dynamic stall on airfoils negatively impacts their aerodynamic performance and can lead to structural damage. Accurate prediction and modelling of the dynamic stall loads are crucial for a more robust design of wings and blades that operate under unsteady conditions susceptible to dynamic stall and for widening the range of operation of these lifting surfaces. Many dynamic stall models rely on empirical parameters that need to be obtained from experimental or numerical data which limits their generalisability. Here, we introduce physically derived time scales to replace the empirical parameters in the Goman-Khrabrov dynamic stall model. The physics-based time constants correspond to the dynamic stall delay and the decay of post-stall load fluctuations. The dynamic stall delay is largely independent of the type of the motion, the Reynolds number, and the airfoil geometry and is described as a function of a normalised instantaneous pitch rate. The post-stall decay is independent of the motion kinematics and is related to the Strouhal number of the post-stall vortex shedding. The general validity of our physics-based time constants is demonstrated using three sets of experimental dynamic stall data covering various airfoil profiles, Reynolds numbers varying from \num{75000} to \num{1000000}, and sinusoidal and ramp-up pitching motions. The use of physics-based time constants generalises the Goman-Khrabrov dynamic stall model and extends its range of application.
\end{abstract}

\section{Modelling dynamic stall}
The phenomenon of dynamic stall refers to the massive flow separation experienced by an airfoil due to an unsteady change in angle of attack beyond the critical static stall angle~\citep{Mcalister1978,McCroskey1981,Carr1988}.
Dynamic stall is associated with a delay in stall onset and stall recovery with respect to static flow separation and boundary layer reattachment.
The delay in stall onset is related to an increase in the maximum attainable lift but when separation occurs, it leads to a rapid aft movement of the centre of pressure, large nose-down pitching moments on the airfoil section, and an increase in torsional loads on the wings~\citep{mcalister1984application}.
The unsteady dynamic stall load response reduces the aerodynamic efficiency, limits operational and performance boundaries, and decreases structural stability of lifting surfaces operating at high angles of attack or under highly unsteady flow conditions.
Accurate prediction of the unsteady load response is key to design efficient measures to eliminate or control massive flow separation and improve the design and operation of wind turbines, unmanned aerial vehicles, rapidly manoeuvring aircraft, and helicopters~\citep{Leishman2002,reich2011transient,uhlig2017modeling,Smith.2020}.

The characteristic features identified within a full dynamic stall life-cycle include attached flow, the emergence and spreading of flow reversal on the airfoil's suction side, the formation of a large-scale coherent structure referred to as the dynamic stall vortex, the separation of the first dynamic stall vortex indicating stall onset, massive flow separation, and potentially flow reattachment \citep{Carr1977, Shih1992, Mulleners2013}.
Each of these phenomena are characterised by different time scales and governed by different parameters~\citep{Deparday2019, ansell2020multiscale}.
The succession of different flow stages and the interplay between the associated characteristic phenomena and time scales make the modelling and accurate prediction of the dynamic stall load response challenging.

One of the most prominent dynamic stall models in literature, the Beddoes–Leishman model, manages to represent the complexity of the dynamic stall through four interconnected sub-models that predict the aerodynamic coefficients during different phases of dynamic stall~\citep{Leishman1989}.
An attached flow sub-model represents the unsteady but linear aerodynamic forces prior to stall onset as a sum of non-circulatory and circulatory loading.
A separated flow sub-model models the separation of the boundary layer considering the formation and evolution of the leading edge vortex.
It represents the leading-edge and the trailing-edge separation as a function of losses in the circulation and non-linear forces.
A dynamic stall onset model predicts the timing of stall initiation and a dynamic stall vortex model considers the vortex induced aerodynamic forces.
The Beddoes-Leishman model is widely used for helicopter and wind turbine applications and it accurately predicts the dynamic stall load response while offering a physics-based representation of the stall process for incompressible flows.

Various attempts have been made in the past decades to improve or modify individual aspects of the Beddoes-Leishman model.
In a series of papers, \citeauthor{sheng2006new} introduced and improved a new stall onset criteria for low Mach number flows ($\Ma < 0.3$) \citep{sheng2006new,sheng2008modified}.
\citet{hansen2004beddoes} neglected the compressibility effects and the leading edge flow separation of the Beddoes-Leishman model to develop a model that is better suited to wind turbine applications.
\citet{Larsen2007} proposed a computationally cheaper version of the Beddoes-Leishman model by expressing the static lift curve as a function of the fully attached flow lift value and the degree of attachment.
The Beddoes-Leishman model and derivatives provide a detailed representation of the flow physics and predict the dynamic stall load response but they typically rely on more than ten empirical parameters that need to be extracted and tuned based on experiments or numerical simulations.

For active or closed loop flow control applications, the level of detail provided by Beddoes-Leishman type stall models might not be necessary and state-space models with less empirical parameters and reduced complexity are more desirable.
Data-driven approaches combined with system identification techniques are also better suited for modern control application as they allow for a broad range of inputs conditions, and can be designed to deal well with aggressive pitch angle manoeuvres~\citep{brunton2014state,hemati2017parameter,le2018sindy}.
However, most purely data-driven models do not provide much physical insight into the dynamic stall process.
The state-space Goman-Khrabrov model~\citep{Goman1994} is a first order model with two empirical parameters that provides a good compromise between interpretability and simplicity.
The model has received renewed interest in the past decade for active control applications \citep{Williams2015, williams2019feed,  Sedky.202056i, an2021lift}.
It accurately models dynamic stall load hysteresis over a wide range of reduced frequencies and flow conditions where dynamic stall is charactised by a progressive increase of trailing-edge separation.
This is the case for most occurences of dynamic stall on airfoils at moderate and high Reynolds numbers.
The Goman-Khrabrov model is not applicable to flat plate leading-edge stall cases.

\section{Goman-Khrabrov dynamic stall model}
The Goman-Khrabrov model is a nonlinear state-space model that represents the aerodynamic forces and moments produced by unsteady flows with trailing-edge separation and vortex breakdown on wings at high angles of attack~\citep{Goman1994}.
The model predicts the degree of flow attachment on the upper surface of the airfoil or the position of the vortex breakdown described by a single internal state variable $X$.
The value $X = 1$ represents completely attached flow, and $X = 0$ corresponds to fully separated flow.
The state equation for $X$ is given by:
\begin{equation} \label{eq:GomanX}
\kindex{\tau}{1} \frac{dX(t)}{dt} + X(t) = \kindex{X}{0} (\alpha(t) - \kindex{\tau}{2} \dot{\alpha}(t))\quad.
\end{equation}
Here, $\kindex{X}{0} (\alpha)$ is the location of the separation point as a function of the angle of attack under static stall conditions, $\kindex{\tau}{1}$ is the relaxation time constant related to the relaxation of an unsteady force response to a steady state, $\kindex{\tau}{2}$ corresponds to the stall delay time constant, and $\dot{\alpha}$ indicates the instantaneous pitch rate of the airfoil.
The left hand side of \cref{eq:GomanX} is a first-order linear differential model that describes the evolution of the degree of flow attachment during unsteady manoeuvres of the airfoil.
The right hand side is a nonlinear function that returns the value of $X$ for the different angles of attack based on the static response.
Once $X$ is found, the lift response is obtained using Kirchhoff's law:
\begin{equation}\label{eq:Kirchhoff}
\kindex{C}{l}(\alpha) = \left.\frac{d\kindex{C}{l}}{d\alpha}\right|_{0} \sin \alpha \left(\frac{1 + \sqrt{X}}{2}\right)^2,
\end{equation}
where $\left.\dfrac{d\kindex{C}{l}}{d\alpha}\right|_{0}$ is the static lift slope.

The Goman-Khrabrov model includes two empirical parameters $\kindex{\tau}{1}$ and $\kindex{\tau}{2}$ and requires the static evolution of the separation point location with angle of attack $\kindex{X}{0}(\alpha)$ as input.
The separation point location $\kindex{X}{0}(\alpha)$ is typically determined by fitting the Kirchhoff law (\cref{eq:Kirchhoff}) to the static lift polar $\kindex{C}{l}(\alpha)$ of an airfoil at a given Reynolds number.
Once $\kindex{X}{0}(\alpha)$ is obtained, $\kindex{\tau}{1}$ and $\kindex{\tau}{2}$ are empirically determined by applying \cref{eq:GomanX} to unsteady experimental or numerical training data.
The best fit values of $\kindex{\tau}{1}$ and $\kindex{\tau}{2}$ are considered to be constant for a large range of pitching frequencies for a specific airfoil shape \citep{Williams2015, williams2018alleviating, le2018sindy, an2021lift}.

We have applied this procedure to dynamic stall data of a sinusoidally pitching OA209 airfoil at a chord based Reynolds number $\Rey = \num{9.2e5}$ and Mach number $\Ma=\num{0.14}$.
This data has been previously presented in various publications \citep{mulleners2012onset,Mulleners2013,Deparday2019,ansell2020multiscale}.
The data has been obtained in an open test section of a closed loop wind tunnel.
For this study, the data is corrected with the classical open test section corrections specified in \citet{ewald1998wind}.
Pressure based lift coefficient per unit span are obtained by numerically integrating the instantaneously measured chord-wise pressure distributions in the airfoil’s mid-span cross section.
The lift coefficient per unit span is defined as:
$\kindex{C}{l} = L/(1/2 \rho \Uinf^2 c)$, where $L$ is the lift per unit span, $\rho$ is the air density, $\Uinf$ is the wind tunnel velocity, and $c$ is the airfoil chord length.
The static lift polar was obtained by quasi-steadily increasing the angle of attack starting from well below the critical stall angle.

%%%%%%%%%%%%%%%%%%%%%%%%%%%%%%%%%%%%%%%%%%%%%%%%%%%%%%%%%%%%%%%%%%%%%%%%%%%%%%%%
\begin{figure}
\centering
\includegraphics[scale=0.9]{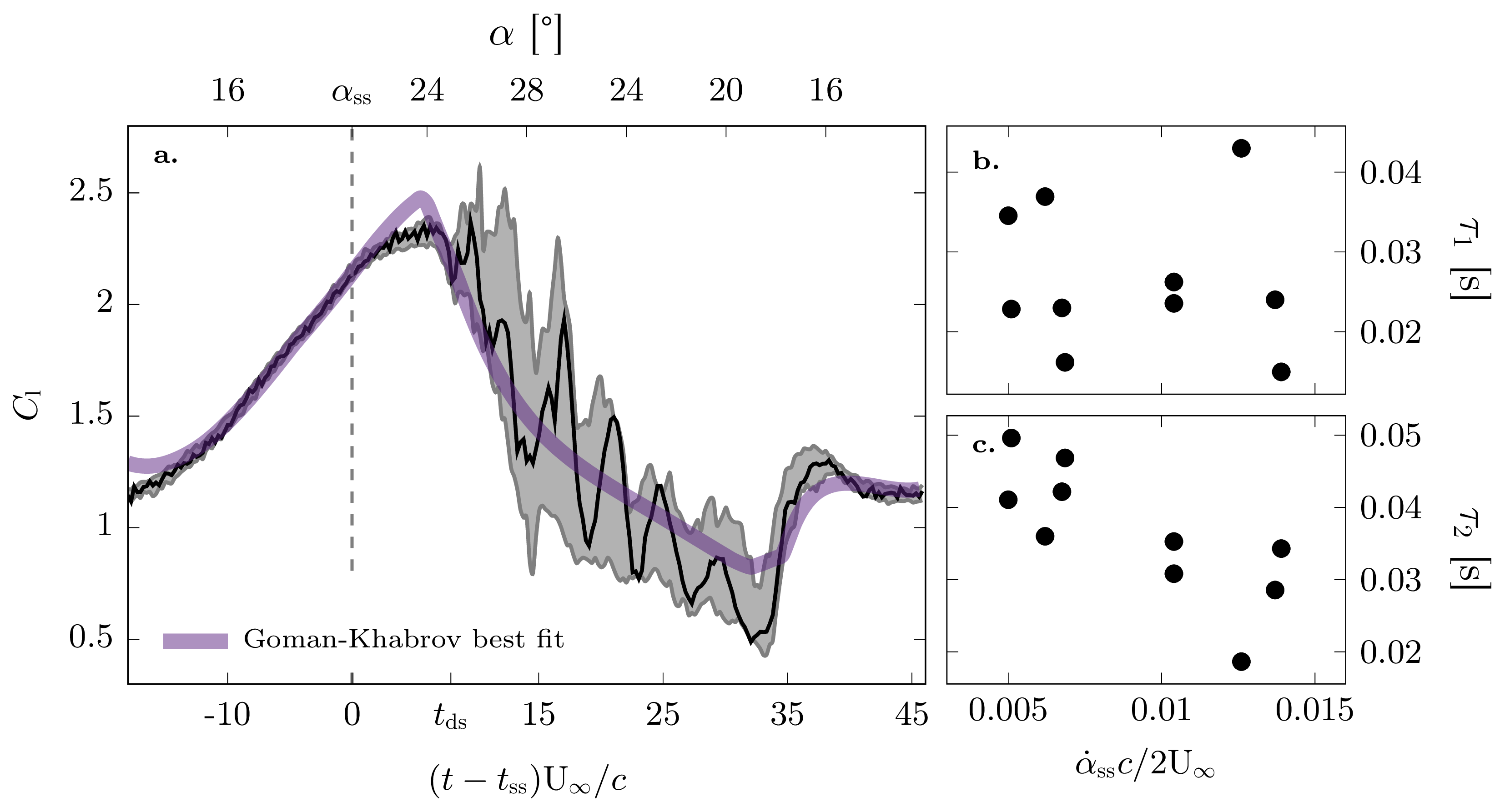}
\caption{\subf{a} Comparison of the evolution of the lift coefficient, $\kindex{C}{l}$, as a function of convective time, $\kindex{t}{c}=t\Uinf/c$, measured experimentally \citep{mulleners2012onset,Mulleners2013} and predicted using the best-fit Goman-Khrabrov model.
The black line represents the measured data for a single cycle of a sinusoidal pitching motion around a mean angle $\kindex{\alpha}{0}=\ang{20}$, with an amplitude $\kindex{\alpha}{1}=\ang{8}$, and a reduced frequency $k=\num{0.05}$.
The grey area indicates the full envelope of the load responses for a total of \num{40} cycle and is a measure for the cycle-to-cycle variations.
Best-fit values of \subf{b} the relaxation time constant \kindex{\tau}{1} and \subf{c} the stall delay time constant \kindex{\tau}{2} for various sinusoidal pitching motions characterised by their normalised effective unsteadiness.}
\label{fig:GK_13_unmod}
\end{figure}
%%%%%%%%%%%%%%%%%%%%%%%%%%%%%%%%%%%%%%%%%%%%%%%%%%%%%%%%%%%%%%%%%%%%%%%%%%%%%%%%

An exemplary result of the Goman-Khrabrov lift coefficient prediction for a sinusoidal pitching motion around a mean angle of attack $\kindex{\alpha}{0}=\ang{20}$, with an amplitude $\kindex{\alpha}{1}=\ang{8}$, and a reduced frequency $k=\num{0.05}$ is presented in \cref{fig:GK_13_unmod}.
The lift coefficient is presented as a function of convective time, with the origin corresponding to the time, $\kindex{t}{ss}$, at which the static stall angle, $\kindex{\alpha}{ss}$, is exceed.
Selected angle of attack values are indicated in the axis on top.
The Goman-Khrabrov prediction is the best fit of the phase averaged lift evolution that was obtained treating $\kindex{\tau}{1}$ and $\kindex{\tau}{2}$ as fitting parameters in \cref{eq:GomanX}.
The model result is compared to the experimentally measured lift evolution in \cref{fig:GK_13_unmod}.
The black line corresponds to the experimental data for a single pitching cycle, the coloured line to the predicted lift coefficient, and the shaded area to the envelope covered by all measured pitching cycles.
The Goman-Khrabrov model predicts well the attached flow regime, the magnitude and timing of the first lift peak, and the general decaying lift trend post dynamic stall, which occurs at $t=\kindex{t}{ds}$.
It is a first-order model that neglects higher-order features in the flow field and cannot reproduce the higher harmonics of the load fluctuations due to subsequent vortex shedding seen in \cref{fig:GK_13_unmod} \citep{Culler2019,an2021lift}.

The values of $\kindex{\tau}{1}$ and $\kindex{\tau}{2}$ that yield the best-fit predictions of the lift evolution are presented in \cref{fig:GK_13_unmod}b,c for various pitching kinematics.
The different pitching kinematics are characterised by their normalised effective unsteadiness $\kindex{\dot{\alpha}}{ss} c/ (2\Uinf)$.
Here, $\kindex{\dot{\alpha}}{ss}$ is the rate of change of the angle of attack when the static stall angle is exceeded.
This instantaneous pitch rate serves as representative effective pitch rate to characterise the time scales of dynamic stall development for sinusoidal motions \citep{mulleners2012onset, Deparday2019, Kissing2020, lefouest2021dynamics}.
The values for $\kindex{\tau}{1}$ that yield the best-fit results, do not show any dependence on the motion unsteadiness and reach an average value of \SI{0.026(9)}{\second} for all pitching motions considered here.
The values for $\kindex{\tau}{2}$ show a general decreases with increasing effective unsteadiness of the driving motion.
In most prominent previous work that uses the Goman-Khrabrov model, the time coefficients have been manually tuned and constant values have been selected for $\kindex{\tau}{1}$ and $\kindex{\tau}{2}$ for a given airfoil profile for all pitching frequencies \citep{Williams2015}.
The empirical selection of the time constants is the main drawback of the Goman-Khrabrov model, which limits it range of applicability.

Following the construction of the Goman-Khrabrov model, the time constants $\kindex{\tau}{1}$ and $\kindex{\tau}{2}$ are physical interpreted as a relaxation time and a stall delay, respectively.
Yet, so far, no solutions have been proposed to replace the empirical parameters in the Goman-Khrabrov dynamic stall model by physically derived time scales.
Here, we aim to close that gap by introducing functional relationships to determine the parameters $\kindex{\tau}{1}$ and $\kindex{\tau}{2}$ based solely on the kinematic input parameters.

\section{Novel physics-based time constants}
We start with the second time constant $\kindex{\tau}{2}$, which is the stall delay constant.
Instead of empirically fixing this constant for a range of reduced frequencies, we will directly connected $\kindex{\tau}{2}$ to the physical dynamic stall delay between the time at which the static stall angle is exceeded and the onset of dynamic stall.
The stall delay constant represents the delay in the dynamic stall angle of attack due to an unsteady pitching motion to:
\begin{align}\label{eq:ads1}
\kindex{\alpha}{ds} = \kindex{\alpha}{ss} + \kindex{\tau}{2}\,\kindex{\dot{\alpha}}{ss} \quad.
\end{align}
The difference between the static and the dynamic stall angle can also be found by integrating the pitching motion during the dynamic stall delay, such that:
\begin{align} \label{eq:tau2gen}
\kindex{\tau}{2} = \frac{1}{\kindex{\dot{\alpha}}{ss}} \int_{\kindex{t}{ss}}^{\kindex{t}{ds}} d \alpha = \frac{1}{\kindex{\dot{\alpha}}{ss}} \int_{\kindex{t}{ss}}^{\kindex{t}{ss}+\kindex{\Delta t}{ds}} d \alpha \quad ,
\end{align}
with $\kindex{\Delta t}{ds}$ the temporal dynamic stall delay.
For constant pitch rate ramp-up motions, \cref{eq:tau2gen} yields:
\begin{align}\label{eq:tau2ramp}
\kindex{\tau}{2}=\kindex{\Delta t}{ds}\quad,
\end{align}
and the Goman-Khrabrov stall delay constant equals the temporal dynamic stall delay \citep{Ericsson1988}.
For sinusoidal pitching motions with effective unsteadiness \kindex{\dot{\alpha}}{ss}, \cref{eq:tau2gen} yields:
\begin{equation}\label{eq:tau2sin}
\kindex{\tau}{2} = \frac{\kindex{2\alpha}{1}}{\kindex{\dot{\alpha}}{ss}}  \sin(\pi f \Delta \kindex{t}{ds}) \cos(\pi f \Delta \kindex{t}{ds})\quad.
\end{equation}
For sinusoidal pitching motions, the Goman-Khrabrov stall delay constant is again a function of the temporal dynamic stall delay and three kinematic input parameters: the pitching amplitude, the pitching frequency, and the effective unsteadiness.

%%%%%%%%%%%%%%%%%%%%%%%%%%%%%%%%%%%%%%%%%%%%%%%%%%%%%%%%%%%%%%%%%%%%%%%%%%%%%%%%
\begin{figure}
\centering
\includegraphics[scale=0.9]{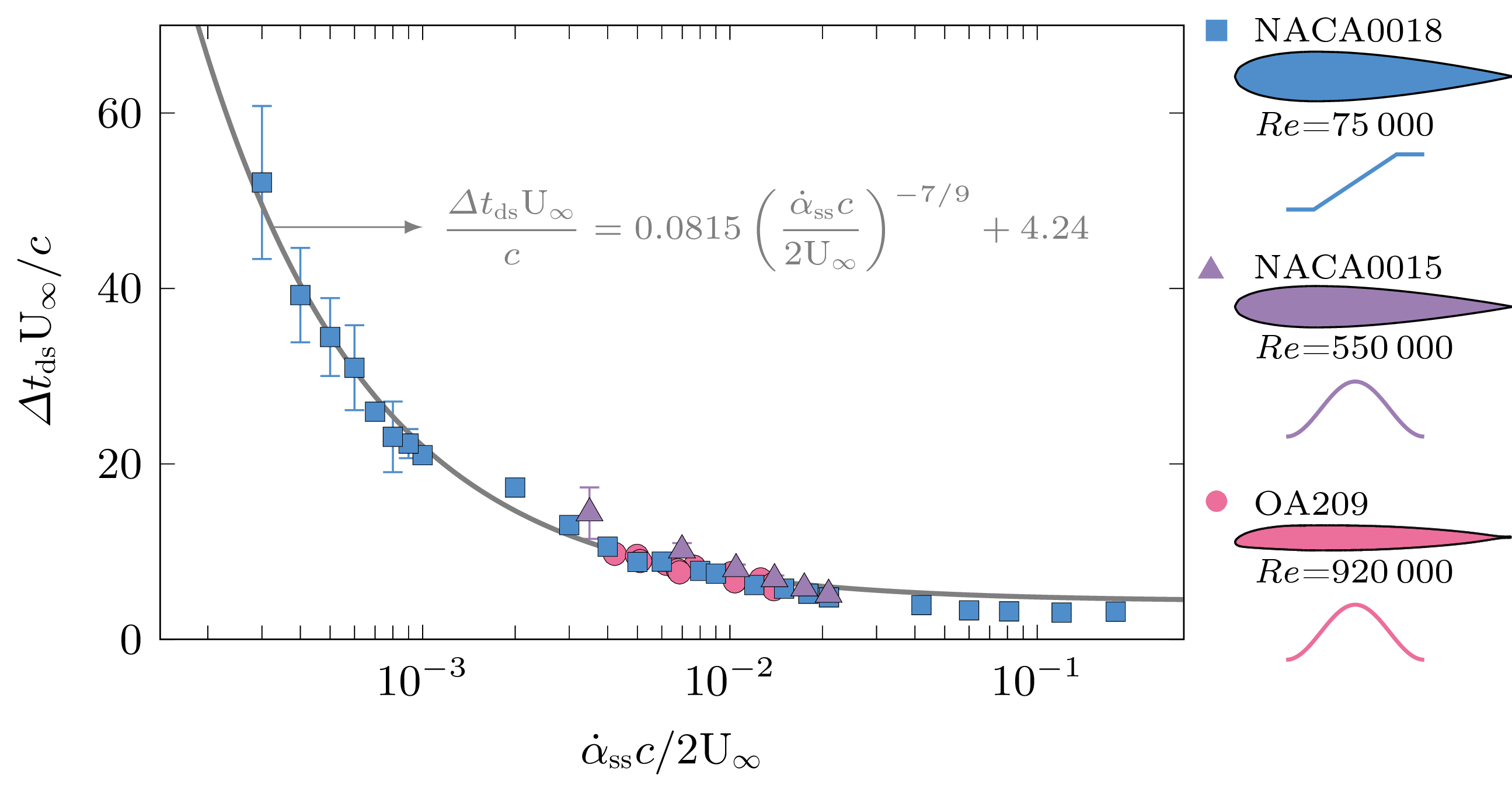}
\caption{Variation of the dynamic stall delay with effective unsteadiness for three different dynamic stall experiments, including linear ramp-up motions of a NACA0015 at \Rey=\num{7.5e4}~\citep{lefouest2021dynamics}, a sinusoidally pitching NACA0015 at \Rey=\num{5.5e5}~\citep{He.2020}, and a sinusoidally pitching 0A209 at \Rey=\num{9.2e5}~\citep{mulleners2012onset}.}
\label{fig:Delta_tds}
\end{figure}
%%%%%%%%%%%%%%%%%%%%%%%%%%%%%%%%%%%%%%%%%%%%%%%%%%%%%%%%%%%%%%%%%%%%%%%%%%%%%%%%

\Cref{eq:tau2ramp,eq:tau2sin} would only truly aid to reduce the empiricism of the Goman-Khrabrov model, if the dynamic stall delay is know.
Recently, \citet{lefouest2021dynamics} compared the stall delays for three different airfoils, a NACA0015, a NACA0018, and an OA209, at different Reynolds numbers ranging from \numrange{7.5e4}{e6}, and covering linear ramp-up and sinusoidal pitching motions.
The data points in \cref{fig:Delta_tds} have different colours and marker shapes referring to the different airfoil shapes, Reynolds number values, and motion kinematics according to the legend on the right of the figure.
All measured dynamic stall delays decrease with increasing motion unsteadiness and converge towards a minimum value for high values of the unsteadiness.
This overall decay of the stall delay is fitted by a power law:
\begin{equation}\label{eq:deltatds}
\kindex{\Delta t}{ds}\Uinf/c = 0.0815 \left (\frac{\kindex{\dot{\alpha}}{ss}c}{2 \Uinf} \right)^{-7/9} + 4.24 \quad,
\end{equation}
for all deep dynamic stall cases considered, regardless of the airfoil geometries, motion kinematics, and Reynolds number (\cref{fig:Delta_tds}).
The R-square value of the fit is \num{0.978}.

The overall stall delay is the sum of the delays attributed to the two stages that are identified in the dynamic stall development process \citep{Mulleners2013, Deparday2019}.
Dynamic stall development includes a reaction stage, which is characterised by accumulation of vorticity in the shear layer, followed by a vortex formation stage, during which the shear layer rolls up into a large scale dynamic stall vortex.
These two stages are characterised by different time responses.
The duration of the first stage decreases with increasing motion unsteadiness.
The duration of the second stage is independent of the motion unsteadiness and indicates the minimum time required for a dynamic stall vortex to form.
The lower limit of the stall reponse $\kindex{\Delta t}{ds} (\kindex{\dot{\alpha}}{ss}\rightarrow\infty)$ corresponds to the duration of the vortex formation stage and is of the order of the general characteristic vortex formation numbers observed for a variety of vortex generators \citep{Dabiri2009}.

Using the universal expression for the dynamic stall delay provided by \cref{eq:deltatds}, we can find $\kindex{\tau}{2}$ solely based on the input parameters.
This reduces the empiricism in the Goman-Khrabrov dynamic stall model and generalises it.\\

The time constant $\kindex{\tau}{1}$ is the relaxation time constant which represents the relaxation time for a perturbed or separated flow to recover to a steady state \citep{Goman1994}.
The post-stall regime of a deep dynamic stall cycle is characterised by repeated large-scale vortex shedding which leads to large fluctuations in the lift.
The post-stall lift fluctuations due to vortex shedding are most clearly visible in the lift evolution of a single pitching cycle represented by the black curve in \cref{fig:GK_13_unmod}.
Increased flow perturbations post-stall, cause fluctuations or jitter in the post-stall vortex formation and shedding between successive pitching cycles.
This leads to the so-called cycle-to-cycle variations in the lift evolution.
The shaded area in \cref{fig:GK_13_unmod} indicates the region between the minimum and maximum values observed across a total of forty pitching cycles.
When the flow is attached, at the beginning of the cycle up to dynamic stall onset at \kindex{t}{ds} and after reattachment, there are little to no cycle-to-cycle variations.
The cycle-to-cycle variations appear post stall and can lead to broadening or even elimination of the post-stall lift fluctuations in the phase-averaged lift cycle.

The Goman-Khrabrov is a first-order model and can not predict the post-stall lift fluctuations.
To provide insight into the characteristic time scales that govern the post-stall lift decay, we present the magnitude of the post-stall lift peaks and their timing for multiple pitching cycles and different sinosoidal pitching motions of the OA209 airfoil at $\Rey = \num{9.2e5}$.
The timing of the lift peaks is presented with respect to the onset of dynamic stall.
The colour of the markers indicates the normalised effective unsteadiness that characterises the individual sinusoidal pitching kinematics ($\kindex{\alpha}{0}\in\{\ang{18},\ang{20},\ang{22}\}$, $\kindex{\alpha}{1}\in\{\ang{6},\ang{8}\}$, $k\in\{0.05,0.075,0.1\}$).

%%%%%%%%%%%%%%%%%%%%%%%%%%%%%%%%%%%%%%%%%%%%%%%%%%%%%%%%%%%%%%%%%%%%%%%%%%%%%%%%
\begin{figure}
\centering
\includegraphics[scale=0.9]{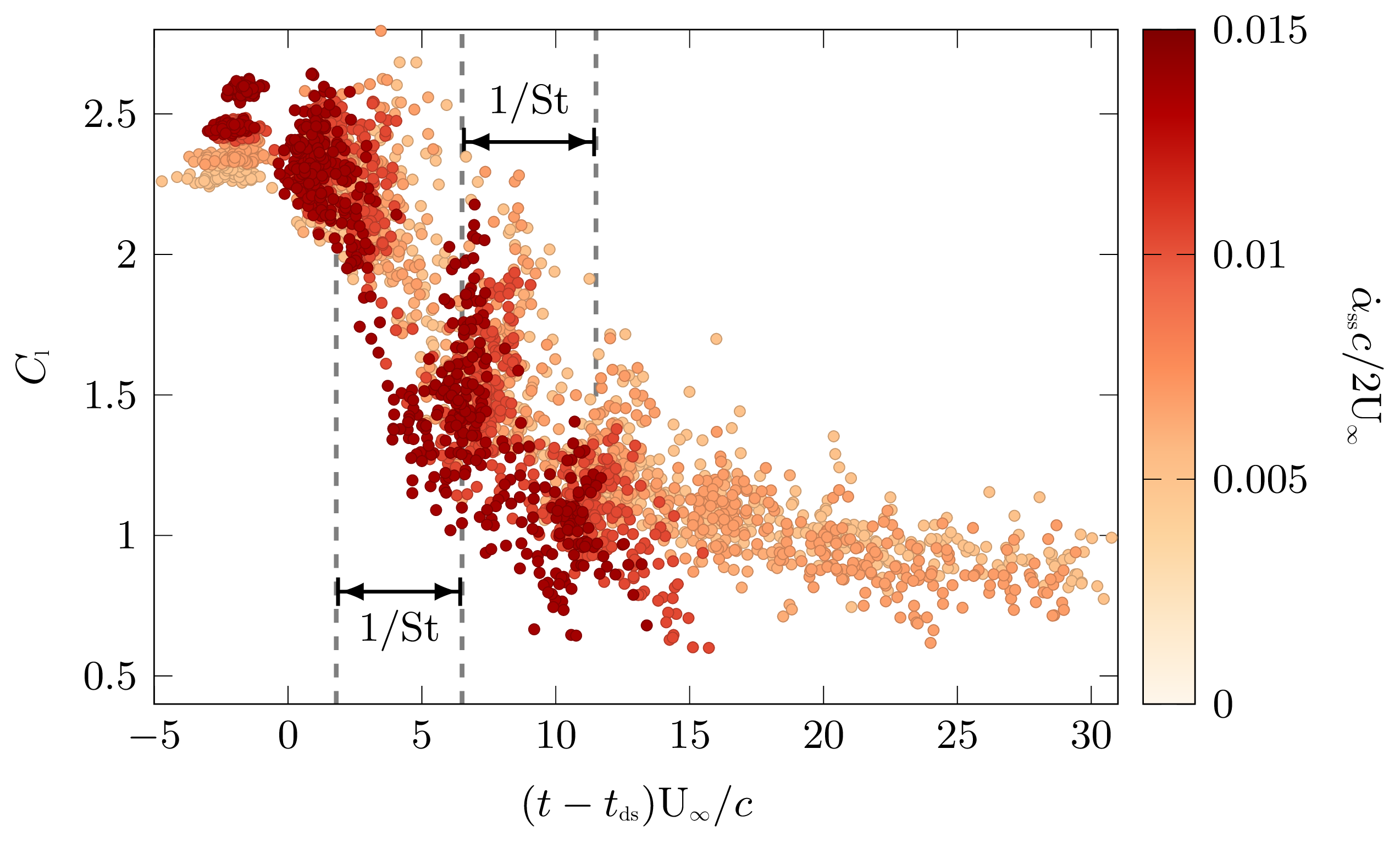}
\caption{Local lift peaks as a function of the convective time post dynamic stall onset for various cycles of different sinusoidal pitching motions for the OA209 airfoil at $\Rey = \num{9.2e5}$.
The colour of the markers refers to the normalised effective unsteadiness of the various motions.}
\label{fig:Lift_pks_all}
\end{figure}
%%%%%%%%%%%%%%%%%%%%%%%%%%%%%%%%%%%%%%%%%%%%%%%%%%%%%%%%%%%%%%%%%%%%%%%%%%%%%%%%

\begin{figure}
\centering
\includegraphics[scale=0.9]{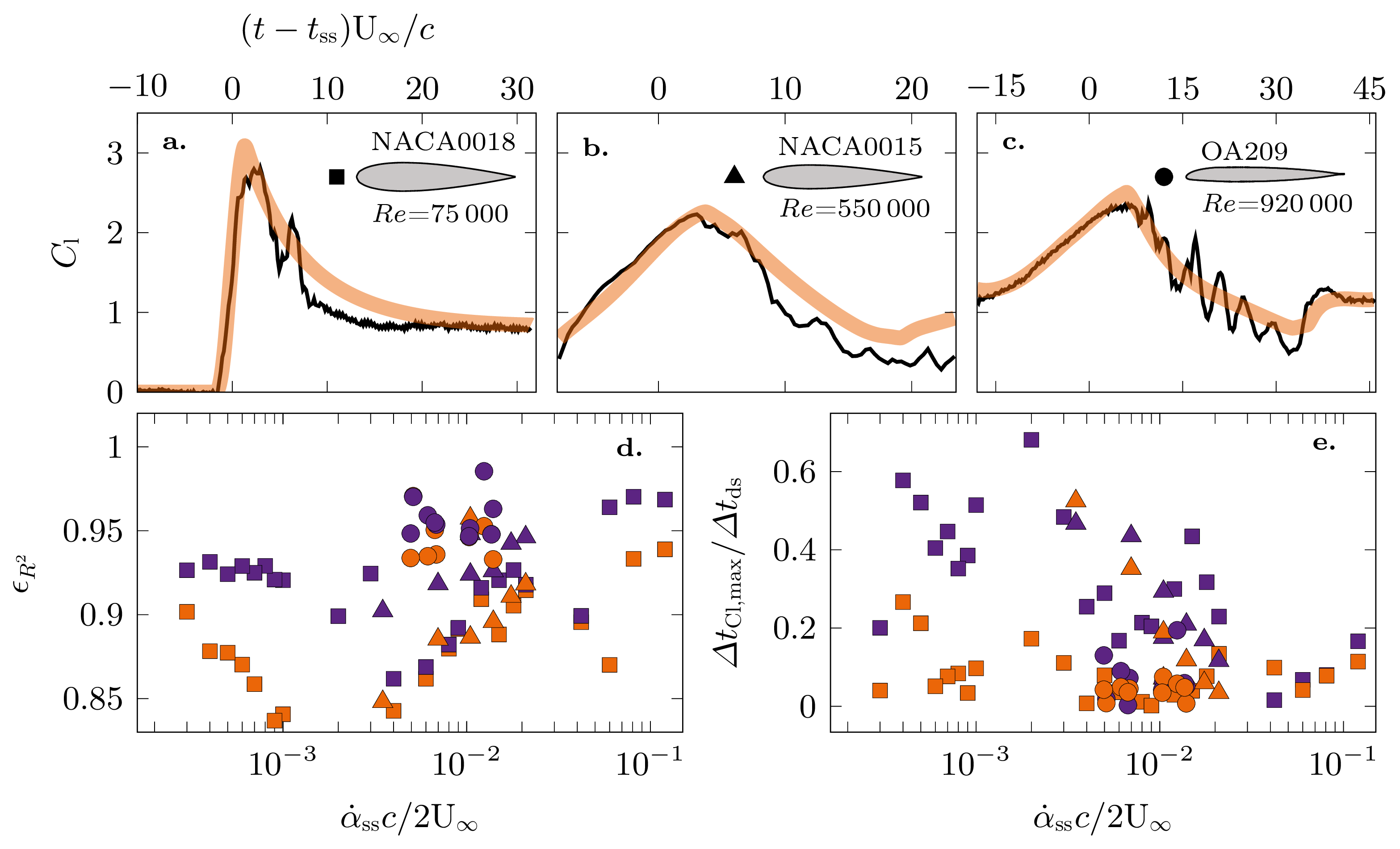}
\caption{Comparison of the temporal evolution of the experimental lift coefficient and the prediction of Goman-Khrabrov model using the physics-based time coefficients for a ramp-up motion and two sinusoidal pitching motions.
($\kindex{\dot{\alpha}}{ss} =$ \subf{a} \num{0.12}, \subf{b} \num{0.014}, \subf{c} \num{0.007}).
\subf{d} $R^2$-error between the experimental lift evolution and its prediction by the Goman-Khrabrov model.
\subf{e} Relative error in the model's prediction of the timing of the first lift peak.
Purple markers indicate results from the best-fit Goman-Khrabrov model and orange markers indicate the results of the model with the novel physics-based time constants.
}
\label{fig:GK_13}
\end{figure}

The lift peaks corresponding to different sinusoidal pitching motions align surprising well despite the large variation in the angle of attack at which they occur.
Independent of the motion unsteadiness and the local angle of attack, the local lift peaks decay with increasing convective time post stall.
The decay has two parts.
The initial decay is steep and covers approximately twelve convective times.
The lift peaks during this initial phase seem to cluster in three groups with a spacing of approximately four to five convective times.
This corresponds to the lower limit of the stall reponse $\kindex{\Delta t}{ds} (\kindex{\dot{\alpha}}{ss}\rightarrow\infty)=4.24\ c/\Uinf$ in \cref{fig:Delta_tds}.
An interval of \num{4.24} convective times is equivalent to a Strouhal number of \num{0.235}.
The decay of the lift peaks becomes less steep but remains more or less linear for longer convective times post stall.
As the post-stall vortex shedding is independent of the motion unsteadiness and governed by convective time scales, we propose to replace the relaxation time constant $\kindex{\tau}{1}$ by the lower limit of the stall reponse or:
\begin{equation}\label{eq:tau1}
\kindex{\tau}{1} = \kindex{\Delta t}{ds} (\kindex{\dot{\alpha}}{ss}\rightarrow\infty)=4.24\, \frac{c}{\Uinf}.
\end{equation}
For the data presented in \cref{fig:GK_13}, this would yield $\kindex{\tau}{1} = \SI{0.0254}{\second}$ which in the middle of the best fit values.\\

To validate the performance of the Goman-Khrabrov model with our physics-based time constants, we have applied the model to the three experimental data sets mentioned above.
The resulting predictions of the Goman-Khrabrov model are compared with a selected ensemble-averaged response for the ramp-up motion and phase-averaged lift responses for the pitching motions in \cref{fig:GK_13}a-c.
The model based on our proposed time scales predicts well the overall evolution of the lift, the timing of the first lift peak, and the general post stall lift decay.
The magnitude of the first lift peak is well predicted for the sinusoidal motions, but under predicted for the ramp-up motion.
Yet, this would be the same for the best fit Goman-Khrabrov prediction.

%%%%%%%%%%%%%%%%%%%%%%%%%%%%%%%%%%%%%%%%%%%%%%%%%%%%%%%%%%%%%%%%%%%%%%%%%%%%%%%%
%%%%%%%%%%%%%%%%%%%%%%%%%%%%%%%%%%%%%%%%%%%%%%%%%%%%%%%%%%%%%%%%%%%%%%%%%%%%%%%%

To qualitatively assess the performance of the newly introduced time constants, we calculated two metrics of interest for the best-fit model and our modified version for all dynamic stall cases within our three experimental data sets.
The first metric is the coefficient of determination, or $R^2$-value.
This metric gives us a general measure of how well the experimental data is replicated by the model.
The second metric is the relative time difference between the timing of the main lift peak in the experimental data and the model prediction.
This is a more local measure that is of particular importance for the prediction of dynamic stall.
The results of both metrics are summarised as a function of the effective unsteadiness, which characterises the motion, in \cref{fig:GK_13}d-e.
Purple markers indicate results from the best-fit Goman-Khrabrov model and orange markers indicate the results of the model with the novel physics-based time constants.
The shapes of the markers refer to the different data sets as indicated in \cref{fig:Delta_tds} and on \cref{fig:GK_13}a-c.

In general, with $R^2$-values above \num{0.85}, both Goman-Khrabrov predictions provide an accurate model of all the experimental data considered here.
The predictions using the time constants based on \cref{eq:tau1,eq:tau2gen} have $R^2$-values that are only marginally lower than the best fit results but they provide a better prediction of the timing of the first lift peak.

\section{Conclusion}
We have generalised the Goman-Khrabrov dynamic stall model and reduced its empiricism by introducing physically derived time scales to replace the empirical parameters in the original model.
The first time constant represents the shedding period of the post-stall load fluctuations.
The post-stall decay rate is independent of the motion kinematics and the time constant is obtained assuming Strouhal number $\St=0.25$.
The second time constant represents the temporal dynamic stall delay.
We presented a generally applicable functional relationship between the non-dimensional stall delay and the normalised instantaneous pitch rate at static stall.
This relationship is largely independent of the type of the motion, the Reynolds number, and the airfoil geometry based on three available dynamic stall data sets.
Using this general expression for the dynamic stall delay, we compute the second time constant directly from the input parameters.
The Goman-Khrabrov model with our newly defined time constants gives excellent predictions of the unsteady lift evolutions for a large set of experimental dynamic stall data covering different airfoil shapes, motion types, a range of Reynolds numbers from \numrange{7.5e4}{e6}, and a large range of reduced frequencies.
The use of our physics-based time constants generalises the Goman-Khrabrov dynamic stall model and opens new opportunities for closed-loop flow control applications.

\section*{Acknowledgments}
The work presented is supported by the SNSF Assistant Professor energy grant number PYAPP2\_173652.
\section*{Declaration of Interests}
The authors report no conflict of interest.
\bibliographystyle{jfm}
\bibliography{ds_model}

\begin{thebibliography}{35}
\expandafter\ifx\csname natexlab\endcsname\relax\def\natexlab#1{#1}\fi
\def\au#1{#1} \def\ed#1{#1} \def\yr#1{#1}\def\at#1{#1}\def\jt#1{\textit{#1}}
  \def\bt#1{#1}\def\bvol#1{\textbf{#1}} \def\vol#1{#1} \def\pg#1{#1}
  \def\publ#1{#1}\def\arxiv#1{#1}\def\org#1{#1}\def\st#1{\textit{#1}}

\bibitem[An {\em et~al.\/}(2021)An, Williams, Eldredge \& Colonius]{an2021lift}
{\sc \au{An, X.}, \au{Williams, D.~R.}, \au{Eldredge, J.~D.} \& \au{Colonius,
  T.}} \yr{2021}  \at{Lift coefficient estimation for a rapidly pitching
  airfoil}.  \jt{Experiments in Fluids}  \bvol{62}~(1),  \pg{1--12}.

\bibitem[Ansell \& Mulleners(2020)]{ansell2020multiscale}
{\sc \au{Ansell, P.~J.} \& \au{Mulleners, K.}} \yr{2020}  \at{Multiscale vortex
  characteristics of dynamic stall from empirical mode decomposition}.
  \jt{AIAA journal}  \bvol{58}~(2),  \pg{600--617}.

\bibitem[Brunton {\em et~al.\/}(2014)Brunton, Dawson \&
  Rowley]{brunton2014state}
{\sc \au{Brunton, S.~L.}, \au{Dawson, S. T.~M.} \& \au{Rowley, C.~W.}}
  \yr{2014}  \at{State-space model identification and feedback control of
  unsteady aerodynamic forces}.  \jt{Journal of Fluids and Structures}
  \bvol{50},  \pg{253--270}.

\bibitem[Carr(1988)]{Carr1988}
{\sc \au{Carr, L.~W.}} \yr{1988}  \at{{Progress in analysis and prediction of
  dynamic stall}}.  \jt{Journal of Aircraft}  \bvol{25}~(1).

\bibitem[Carr {\em et~al.\/}(1977)Carr, McAlister \& McCroskey]{Carr1977}
{\sc \au{Carr, L.~W.}, \au{McAlister, K.~W.} \& \au{McCroskey, W.~J.}}
  \yr{1977}  \bt{{Analysis of the Development of Dynamic Stall Based on
  Oscillating Airfoil Experiments}}. NASA technical note TN D-8382.
  \org{NASA}.

\bibitem[Culler \& Farnsworth(2019)]{Culler2019}
{\sc \au{Culler, E. C.~E.} \& \au{Farnsworth, J. A.~N.}} \yr{2019}  \at{{Pitch
  rate induced separation delay modeling of dynamic stall and stall flutter}}.
  \jt{AIAA Scitech 2019 Forum} ~(January),  \pg{1--10}.

\bibitem[Dabiri(2009)]{Dabiri2009}
{\sc \au{Dabiri, John~O.}} \yr{2009}  \at{{Optimal Vortex Formation as a
  Unifying Principle in Biological Propulsion}}.  \jt{Annual review of fluid
  mechanics}  \bvol{41},  \pg{17---33}.

\bibitem[Deparday \& Mulleners(2019)]{Deparday2019}
{\sc \au{Deparday, J.} \& \au{Mulleners, K.}} \yr{2019}  \at{{Modeling the
  interplay between the shear layer and leading edge suction during dynamic
  stall}}.  \jt{Physics of Fluids}  \bvol{31}~(10),  \pg{107104}.

\bibitem[Ericsson \& Reding(1988)]{Ericsson1988}
{\sc \au{Ericsson, L.~E.} \& \au{Reding, J.~P.}} \yr{1988}  \bt{{Fluid
  Mechanics of Dynamic Stall Part I. Unsteady Flow Concepts}}.  \org{{\em Tech.
  Rep.\/}}.

\bibitem[Ewald(1998)]{ewald1998wind}
{\sc \au{Ewald, B.~F.}} \yr{1998}  \bt{Wind tunnel wall corrections (la
  correction des effets de paroi en soufflerie)}. {\em Tech. Rep.\/}.
  \org{Advisory Group for Aerospace Research and Development Neuilly-sur-Seine
  (France)}.

\bibitem[Goman \& Khrabrov(1994)]{Goman1994}
{\sc \au{Goman, M.} \& \au{Khrabrov, A.}} \yr{1994}  \at{{State-space
  representation of aerodynamic characteristics of an aircraft at high angles
  of attack}}.  \jt{Journal of Aircraft}  \bvol{31}~(5).

\bibitem[Hansen {\em et~al.\/}(2004)Hansen, Gaunaa \&
  Madsen]{hansen2004beddoes}
{\sc \au{Hansen, M.~H.}, \au{Gaunaa, M.} \& \au{Madsen, H.~A.}} \yr{2004}
  \at{A beddoes-leishman type dynamic stall model in state-space and indicial
  formulations} .

\bibitem[He {\em et~al.\/}(2020)He, Deparday, Siegel, Henning \&
  Mulleners]{He.2020}
{\sc \au{He, G.}, \au{Deparday, J.}, \au{Siegel, L.}, \au{Henning, A.} \&
  \au{Mulleners, K.}} \yr{2020}  \at{{Stall Delay and Leading-Edge Suction for
  a Pitching Airfoil with Trailing-Edge Flap}}.  \jt{AIAA Journal}
  \bvol{58}~(12),  \pg{5146--5155}.

\bibitem[Hemati {\em et~al.\/}(2017)Hemati, Dawson \&
  Rowley]{hemati2017parameter}
{\sc \au{Hemati, M.~S.}, \au{Dawson, S. T.~M.} \& \au{Rowley, C.~W.}} \yr{2017}
   \at{Parameter-varying aerodynamics models for aggressive pitching-response
  prediction}.  \jt{AIAA journal}  \bvol{55}~(3),  \pg{693--701}.

\bibitem[Kissing {\em et~al.\/}(2020)Kissing, Kriegseis, Li, Feng, Hussong \&
  Tropea]{Kissing2020}
{\sc \au{Kissing, J.}, \au{Kriegseis, J.}, \au{Li, Z.}, \au{Feng, L.},
  \au{Hussong, J.} \& \au{Tropea, C.}} \yr{2020}  \at{{Insights into leading
  edge vortex formation and detachment on a pitching and plunging flat plate}}.
   \jt{Experiments in Fluids}  \bvol{61}~(9),  \pg{208}.

\bibitem[Larsen {\em et~al.\/}(2007)Larsen, Nielsen \& Krenk]{Larsen2007}
{\sc \au{Larsen, J.~W.}, \au{Nielsen, S.~R.K.} \& \au{Krenk, S.}} \yr{2007}
  \at{{Dynamic stall model for wind turbine airfoils}}.  \jt{Journal of Fluids
  and Structures}  \bvol{23}~(7).

\bibitem[Le~Fouest {\em et~al.\/}(2021)Le~Fouest, Deparday \&
  Mulleners]{lefouest2021dynamics}
{\sc \au{Le~Fouest, S.}, \au{Deparday, J.} \& \au{Mulleners, K.}} \yr{2021}
  \at{The dynamics and timescales of static stall}.  \jt{arXiv preprint
  arXiv:2102.04485} .

\bibitem[Le~Provost {\em et~al.\/}(2018)Le~Provost, Williams \&
  Brunton]{le2018sindy}
{\sc \au{Le~Provost, M.}, \au{Williams, D.~R.} \& \au{Brunton, S.}} \yr{2018}
  Sindy analysis of disturbance and plant model superposition on a rolling
  delta wing.  \bt{In {\em AIAA Flow control conference\/}}.

\bibitem[Leishman(2002)]{Leishman2002}
{\sc \au{Leishman, J.~G.}} \yr{2002}  \at{{Challenges in modelling the unsteady
  aerodynamics of wind turbines}}.  \jt{Wind Energy}  \bvol{5}~(2-3),  \pg{85
  132}.

\bibitem[Leishman \& Beddoes(1989)]{Leishman1989}
{\sc \au{Leishman, J.~G.} \& \au{Beddoes, T.~S.}} \yr{1989}  \at{{A
  semi-empirical model for dynamic stall}}.  \jt{Journal of the American
  Helicopter Society}  \bvol{34}~(3).

\bibitem[Mcalister {\em et~al.\/}(1978)Mcalister, Carr \&
  Mccroskey]{Mcalister1978}
{\sc \au{Mcalister, K.~W.}, \au{Carr, L.~W.} \& \au{Mccroskey, W.~J.}}
  \yr{1978}  \at{{Dynamic stall experiments on the NACA 0012 airfoil}}.
  \jt{Journal of Applied Physics}  \bvol{107}~(9).

\bibitem[McAlister {\em et~al.\/}(1984)McAlister, Lambert \&
  Petot]{mcalister1984application}
{\sc \au{McAlister, K.~W.}, \au{Lambert, O.} \& \au{Petot, D.}} \yr{1984}
  \bt{Application of the onera model of dynamic stall}. {\em Tech. Rep.\/}.
  \org{NASA-TP-2399}.

\bibitem[McCroskey(1981)]{McCroskey1981}
{\sc \au{McCroskey, W.~J.}} \yr{1981}  \bt{{The Phenomenon of Dynamic Stall}}.
  \org{{\em Tech. Rep.\/}}.

\bibitem[Mulleners \& Raffel(2012)]{mulleners2012onset}
{\sc \au{Mulleners, K.} \& \au{Raffel, M.}} \yr{2012}  \at{The onset of dynamic
  stall revisited}.  \jt{Experiments in fluids}  \bvol{52}~(3),  \pg{779--793}.

\bibitem[Mulleners \& Raffel(2013)]{Mulleners2013}
{\sc \au{Mulleners, K.} \& \au{Raffel, M.}} \yr{2013}  \at{{Dynamic stall
  development}}.  \jt{Experiments in Fluids}  \bvol{54}~(2).

\bibitem[Reich {\em et~al.\/}(2011)Reich, Eastep, Altman \&
  Albertani]{reich2011transient}
{\sc \au{Reich, G.~W.}, \au{Eastep, F.~E.}, \au{Altman, A.} \& \au{Albertani,
  R.}} \yr{2011}  \at{Transient poststall aerodynamic modeling for extreme
  maneuvers in micro air vehicles}.  \jt{Journal of Aircraft}  \bvol{48}~(2),
  \pg{403--411}.

\bibitem[Sedky {\em et~al.\/}(2020)Sedky, Jones \& Lagor]{Sedky.202056i}
{\sc \au{Sedky, G.}, \au{Jones, A.~R.} \& \au{Lagor, F.~D.}} \yr{2020}
  \at{{Lift Regulation During Transverse Gust Encounters Using a Modified
  Goman–Khrabrov Model}}.  \jt{AIAA Journal}  \bvol{58}~(9),  \pg{1--11}.

\bibitem[Sheng {\em et~al.\/}(2008)Sheng, Galbraith \&
  Coton]{sheng2008modified}
{\sc \au{Sheng, W.}, \au{Galbraith, R.~A.} \& \au{Coton, F.~N.}} \yr{2008}
  \at{A modified dynamic stall model for low mach numbers}.  \jt{Journal of
  Solar Energy Engineering}  \bvol{130}~(3).

\bibitem[Sheng {\em et~al.\/}(2006)Sheng, Galbraith \& Coton]{sheng2006new}
{\sc \au{Sheng, W.}, \au{Galbraith, R. A.~M.} \& \au{Coton, F.~N.}} \yr{2006}
  \at{A new stall-onset criterion for low speed dynamic-stall}.  \jt{Journal of
  Solar Energy Engineering}  \bvol{128}~(4),  \pg{461--471}.

\bibitem[Shih {\em et~al.\/}(1992)Shih, Lourenco, Dommelen \&
  Krothapalli]{Shih1992}
{\sc \au{Shih, C}, \au{Lourenco, L}, \au{Dommelen, L~Van} \& \au{Krothapalli,
  A}} \yr{1992}  \at{{Unsteady Flow Past an Airfoil Pitching at a Constant
  Rate}}.  \jt{AIAA Journal}  \bvol{30}~(5),  \pg{1153--1161}.

\bibitem[Smith {\em et~al.\/}(2020)Smith, Jones, Ayancik, Mulleners \&
  Naughton]{Smith.2020}
{\sc \au{Smith, M.~J.}, \au{Jones, A.~R.}, \au{Ayancik, F.}, \au{Mulleners, K.}
  \& \au{Naughton, J.~W.}} \yr{2020} {An Assessment of the State-of-the-Art
  from the 2019 ARO Dynamic Stall Workshop}.  \bt{In {\em AIAA Aviation
  forum\/}}. virtual.

\bibitem[Uhlig \& Selig(2017)]{uhlig2017modeling}
{\sc \au{Uhlig, D.~V.} \& \au{Selig, M.~S.}} \yr{2017}  \at{Modeling micro air
  vehicle aerodynamics in unsteady high angle-of-attack flight}.  \jt{Journal
  of Aircraft}  \bvol{54}~(3),  \pg{1064--1075}.

\bibitem[Williams {\em et~al.\/}(2015)Williams, An, Iliev, King \&
  Rei{\ss}ner]{Williams2015}
{\sc \au{Williams, D.~R.}, \au{An, X.}, \au{Iliev, S.}, \au{King, R.} \&
  \au{Rei{\ss}ner, F.}} \yr{2015}  \at{{Dynamic hysteresis control of lift on a
  pitching wing}}.  \jt{Experiments in Fluids}  \bvol{56}~(5).

\bibitem[Williams {\em et~al.\/}(2019)Williams, Greenblatt, M{\"u}ller-Vahl,
  Santra \& Rei{\ss}ner]{williams2019feed}
{\sc \au{Williams, D.~R.}, \au{Greenblatt, D.}, \au{M{\"u}ller-Vahl, H.},
  \au{Santra, S.} \& \au{Rei{\ss}ner, F.}} \yr{2019}  \at{Feed-forward dynamic
  stall control model}.  \jt{AIAA Journal}  \bvol{57}~(2),  \pg{608--615}.

\bibitem[Williams \& King(2018)]{williams2018alleviating}
{\sc \au{Williams, D.~R.} \& \au{King, R.}} \yr{2018}  \at{Alleviating unsteady
  aerodynamic loads with closed-loop flow control}.  \jt{AIAA Journal}
  \bvol{56}~(6),  \pg{2194--2207}.

\end{thebibliography}

\end{document}